\documentclass[preprint, superscriptaddress]{revtex4-2}
\usepackage{graphicx}
\usepackage{physics}

\renewcommand{\a}{\alpha}

\DeclareMathOperator*{\argmax}{arg\,max}

\begin{document}

\title{Entanglement-based quantum key distribution with non-Gaussian continuous variables}

\author{Hao Jeng}
\email{hao.jeng@mpinat.mpg.de}
\affiliation{Centre for Quantum Computation and Communication Technology, Department of Quantum Science, Australian National University, Canberra, ACT 2601, Australia}
\affiliation{Department of Ultrafast Dynamics, Max Planck Institute for Multidisciplinary Sciences, Göttingen, Germany}
\affiliation{IV. Physical Institute, University of Göttingen, Göttingen, Germany}

\author{Ping Koy Lam}
\affiliation{Centre for Quantum Computation and Communication Technology, Department of Quantum Science, Australian National University, Canberra, ACT 2601, Australia}
\affiliation{Quantum Innovation Centre (Q.InC), Agency for Science Technology and Research (A*STAR), 2 Fusionopolis Way, Innovis \#08-03, Singapore 138634, Republic of Singapore}

\author{Syed M. Assad}
\email{cqtsma@gmail.com}
\affiliation{Centre for Quantum Computation and Communication Technology, Department of Quantum Science, Australian National University, Canberra, ACT 2601, Australia}
\affiliation{Quantum Innovation Centre (Q.InC), Agency for Science Technology and Research (A*STAR), 2 Fusionopolis Way, Innovis \#08-03, Singapore 138634, Republic of Singapore}






\begin{abstract}
Addition of single photons to two-mode-squeezed-vacuum states has the effect of distilling quantum entanglement, and, when deployed in quantum key distribution, should lead also to an increase in the secret key rate. However, the extraction of secret keys from non-Gaussian entangled states is a complex issue and is at present not fully understood. In this paper we describe a technique for adding photons to entangled states, and demonstrate how it leads to an increase in secret key rates and the maximal distance for which keys can be distributed assuming asymptotic conditions. The quantum correlations thus produced were found to be of a highly non-Gaussian character, such that the Gaussian extremity principle returns a negative keyrate and effectively kills the protocol; we have therefore developed methods of analysis that do not require prior assumptions about the state. Although it could have been that the addition of single photons would make the system more fragile, this turned out not to be the case. Rather, the addition of a single photon was found to protect the protocol against both passive and active decoherence.
\end{abstract}

\maketitle


\section{Introduction}

This paper is the second in a series of studies on the action of photon addition on quantum states, first made public in the doctoral thesis of one of us \cite{jeng_thesis}. The first study concerns addition of photons to coherent states, which we observed to bear remarkable similarities with strong cubic phase shifts acting on the vacuum state \cite{jeng} and which are of great importance for universal quantum computation. In this paper, we are interested in the non-Gaussian quantum correlations generated by photon addition acting on two-mode-squeezed-vacuum states, and the question of how secret keys could be extracted from such highly non-Gaussian entangled states.

Quantum key distribution with continuous variables has traditionally focused on Gaussian states, for the reason that it would be much easier to implement in the laboratory. Well-known examples include protocols based on coherent states \cite{grosshans1,grosshans2}, squeezed states \cite{hillary, cerf}, and two-mode-squeezed vacuum states \cite{ralph}. While there are several interesting cases that consider non-Gaussian states arising from coherent modulations that are not normally distributed \cite{leverrier, li}, the security analyses of these schemes rely on the so-called "Gaussian extremity principle" \cite{wolf}, which states that the keyrate corresponding to the Gaussian state with the same covariance matrix serves as a lower bound to the optimal value. This is not an issue if the modulations are very small as the quantum state will very closely resemble a Gaussian distribution, but in general that need not be the case.

Non-Gaussian states are very important to quantum communication. Entanglement distillation of Gaussian states using only Gaussian operations and Gaussian measurements is prohibited by no-go theorems \cite{eisert,fiurasek,giedke}, thus a quantum repeater could not be built using only these elements. Quantum error correction is also forbidden for similar reasons \cite{niset}, so we could not simply correct for transmission errors either. While it is nevertheless possible to construct large-scale networks using trusted-node repeaters, this method of communication exposes the user to a great number of security risks.

When analysing quantum key distribution with complex non-Gaussian states, application of the Gaussian extremity principle is inappropriate on at least two counts. First, the true keyrate could be much higher, as it is only a lower bound. Secondly, the covariance matrix can correspond to a mixed Gaussian state even when the original non-Gaussian state is pure. The apparent decoherence will then be mistakenly attributed to the presence of an eavesdropper, effectively shortening the distance across which quantum key distribution could be established. In the case of photon-added entangled states, one is led to the rather paradoxical result where the entanglement is increased and yet the key distribution protocol becomes compromised.

We wish to attack this problem by analysing the mutual information and the eavesdropper's information in such a way as to account for the non-Gaussian properties of the state. This is possible since the formulae for neither of them (the latter being Holevo's bound \cite{holevo}) fundamentally require the assumption of Gaussian states, which serves only to simplify calculations and give closed form expressions. The keyrate can thus be evaluated numerically in the general case of arbitrary quantum states.

When such an analysis is carried out, we find the behaviour of the keyrate to be perfectly satisfactory: it increases with the number of added photons, along with the degree of entanglement of the state. We shall also see exactly what went wrong with the Gaussian assumption in the case of photon addition---while there is only a slight difference between the mutual information of Gaussian and non-Gaussian states, the eavesdropper's information skyrockets with even just a single photon added and immediately drives the keyrate negative. This is consistent with the idea that non-Gaussian effects gives the illusory appearance of an adversary when viewed through the covariance matrix.

The paper can be divided into two parts. In the first part, we will discuss the method of adding photons to entangled states and the problem of tomographic reconstruction for quantum states with two modes, and use it to study the properties of these states. In the second part, we will analyse the problem of distributing secret keys using these states. Because we are interested in the ideal behaviour of such entangled states in the context of quantum key distribution, throughout this paper we assume asymptotic conditions when obtaining keyrates. 

\section{Photon addition}

The photon creation and annihilation operators, $a^\dagger$ and $a$, offer simple and accessible ways of transforming Gaussian states into non-Gaussian ones. Also known colloquially as photon addition and subtraction, they can be implemented respectively through the conditional absorption or emission of a single photon, viz.,
\begin{align}
    \bra{1}_2 BS(R) \ket{\psi}_1\ket{0}_2 & \approx i \sin^{-1}{R} ~a\ket{\psi},\\
    \bra{0}_2 BS(R) \ket{\psi}_1\ket{1}_2 & \approx i \sin^{-1}{R} ~a^\dagger \ket{\psi},
\end{align}
where $BS(R)$ denotes the beamsplitter unitary with reflectivity $R$, with the understanding that $R$ is very small. In the case of photon addition, a popular implementation is through injection of the state $\ket{\psi}$ into one mode of a pair of spontaneous down-converted beams and heralding on detecting a single photon in the other mode \cite{zavatta}, which would be easier than conditioning on the detection of nothing.

Of the two operations, photon subtraction is easier to implement and has been studied in greater detail. Arguably, one of the most important results is the transformation of squeezed states into cat states through single photon subtraction \cite{ourjoumtsev, neergaard-nielsen, wakui}. Although photon addition is harder to implement and its actions understood to a lesser extent, we have shown quite recently that it can be used to induce cubic phase shifts when acting on coherent states \cite{jeng, fadrny}:
\begin{equation}
    a^\dagger \ket{\alpha} \approx \exp(-i\gamma (a+a^\dagger)^3) \ket{0}
\end{equation}
up to Gaussian operations, with these states being very important as a resource for universal quantum computation \cite{lloyd}. Taken together, photon addition and subtraction allow us to make some the most important quantum states in quantum optics.

When acting on entangled states, the effect of photon addition and subtraction is to distil quantum entanglement \cite{navarrete-benlloch}. For a two-mode-squeezed-vacuum state
\begin{equation}
    \ket{\psi} = \sqrt{1-\lambda^2}\sum_{n=0}^\infty \lambda^n \ket{n}_1\ket{n}_2,
\end{equation}
adding one photon to the first mode we obtain,
\begin{equation}
    a_1^\dagger\ket{\psi} = (1-\lambda^2) \sum_{n=0}^\infty \lambda^n \sqrt{n+1}\ket{n+1}_1 \ket{n}_2,
\end{equation}
which, as it turns out, is also equivalent to photon subtraction on the second mode. This state has stronger number correlations because the coefficients are now elevated by a factor of $\sqrt{n+1}$. When multiple photons are added, the correlations become even stronger,
\begin{equation}
    (a_1^\dagger)^k \ket{\psi} = (1-\lambda^2)^{\frac{k+1}{2}} \sum_{n=0}^\infty \lambda^n \sqrt{{n+k \choose n}} \ket{n+k}_1 \ket{n}_2,
\end{equation}
although in general the state will exhibit much greater deviations from a Gaussian state. Photon addition and its variations \cite{takahashi, kurochkin, ulanov, ourjoumtsev2} illustrate the principle behind entanglement distillation for Gaussian states, which is to transform the state by some process involving a combination of single photon sources and detectors such that the degree of entanglement is increased. In general, it is necessary that this transformation takes the state into a non-Gaussian one, because no-go theorems forbid entanglement distillation of Gaussian states with Gaussian operations \cite{eisert,fiurasek,giedke}.

\section{A post-selective implementation of photon addition}

Analysis of quantum key distribution with non-Gaussian states was carried out on archived data \cite{chrzanowski} that was recorded on an experimental setup where two-mode-squeezed-vacuum states were generated by combining a pair of squeezed states in orthogonal quadratures. The entangled states were exposed to various amounts of optical losses through a beamsplitter on one of its modes (0\%, 25\%, 50\%, 75\%, 90\%, 95\%, 98\%, 99\%), with that mode being subjected to heterodyne detection and the other mode to homodyne detection. The homodyne measurement outcomes were evenly divided between amplitude quadrature measurements and phase quadrature measurements.

The implementation of heterodyne detection on one of the modes makes possible also the implementation of photon addition through a postprocessing technique, without the need for physical single photon ancillary states. Given a general state $\rho$, a single photon addition transforms it as $a^\dagger \rho a$, but if we make heterodyne measurements on this state we find that, 
\begin{equation}
    \bra{\a} a^\dagger \rho a \ket{\a} = |\a|^2 \bra{\a} \rho \ket{\a}, 
\end{equation}
since coherent states are eigenstates of the annihilation operator. This equation expresses the fact that adding a photon before making heterodyne detection is indistinguishable from making the detection and then postselecting the outcomes according to the weighting $|\a|^2$. This is not limited to addition of a single photon, but works in general for an arbitrary number of photons,
\begin{equation}
    \bra{\a} (a^\dagger)^k \rho a^k \ket{\a} = |\a|^{2k} \bra{\a} \rho \ket{\a}, 
\end{equation}
with the only technical limitation being the falling probability of success. As a practical matter, the function $|\a|^{2n}$ is ill-suited as an acceptance probability since for a wide range of outcomes its value is greater than one, and therefore in actuality we use a function that is proportional to this but capped at the value of one:
\begin{equation}
  f(\a)
  \begin{cases}
       \frac{|\a|^{2k}}{|\a_c|^{2k}} &\quad\text{if }  |\a| \le |\a_c|,\\
       1 &\quad\text{if }  |\a| > |\a_c|.\\
     \end{cases}
\end{equation}
All the outcomes that fall outside the cut-off radius $|\a_c|$ are accepted, and as long as a negligible number of outcomes reside in that region, the effects of the procedure will be indistinguishable from actual addition of single photons.

While this expression is somewhat remarkable in the sense that an important non-Gaussian operation can be implemented straightforwardly by postselection, it is not arbitrary. For instance, we cannot directly implement a photon subtraction in front of the heterodyne detection since coherent states are not eigenstates of the photon creation operator, and we also cannot change the heterodyne detection to other measurements. It is the rather unique combination of heterodyne detection and photon addition that makes this possible. The noiseless linear amplification is another example of such an operator that also effectively commutes with heterodyne detection \cite{fiurasek2, walk}, the difference being that the system remains in a Gaussian state. Some authors \cite{li} have worked with a variation of this method that performs instead photon subtraction on the other mode not subjected to the heterodyne detection, but owing to the specific properties of two-mode-squeezed-vacuum states, the schemes are conceptually the same.

Supposing now that we are left with a subsequence of measurements $\{ \a_j, x_j(\theta_j) \}$ after the postselection, where $\a_j$ denotes the heterodyne measurement outcome and $x_j$ and $\theta_j$ denotes the homodyne measurement outcome and measurement angles, then the effective quantum state can be reconstructed through maximum likelihood estimation \cite{lvovsky}. By forming the projection operators 
$\pi_j = \ket{\a_j} \otimes \ket{x_j(\theta_j)} \bra{\a_j} \otimes \bra{x_j(\theta_j)}$, we can calculate the operator
\begin{equation}
    R = \sum_j \frac{\pi_j}{\Tr(\pi_j \rho)}
\end{equation}
and obtain the state that corresponds with the highest probability to our observations through fixed point iteration $\rho \rightarrow R \rho R$, normalizing the density matrix to unit trace at each step. The properties of the operator $R$ and the normalization ensures that the density matrix correspond to a perfectly physical quantum state after every iteration.

The reconstruction algorithm depends on two parameters: the photon number truncation for the density matrices and the filter cut-off $|\a_c|$. Due to the size of Hilbert spaces for entangled states (the size of two-mode density matrices scaling as the quartic power of the photon number truncation), it is necessary to choose these parameters carefully in order to have an accurate reconstruction that could still be completed within a reasonable amount of time. After much trial and error, we eventually settled on a photon number truncation of 10 (i.e. Fock states of up to ten photons are kept) and a filter cut-off of $|\a_c|^2 = 6$. The photon number truncation is chosen so that the density matrix is large enough to accommodate up to three photon addition accurately, which is roughly the maximum number of photons the probability of success will tolerate. The filter cut-off is chosen so as to minimize unphysical non-Gaussian artefacts. that arise from imposing a postselection filter that is non-differentiable on a smoothly varying heterodyne measurement distribution. With this filter cut-off, at least 99.99\% of data reside within the cut-off radius. These settings are such that we have left over one million data points for each homodyne quadrature, which is enough to reconstruct the state accurately. Finally, in order to save time, we manually split the task of running the reconstruction for various datasets over five different computing servers, each with at least 10 physical cores and 32 GB of RAM. The great degree of optimisation and parallelisation enabled the reconstructions to be completed within approximately two days, otherwise we estimate that it could have taken between weeks to a month.

\section{Reconstruction of the non-Gaussian state}

\begin{figure}[!h]
\centering\includegraphics[width=0.75\textwidth]{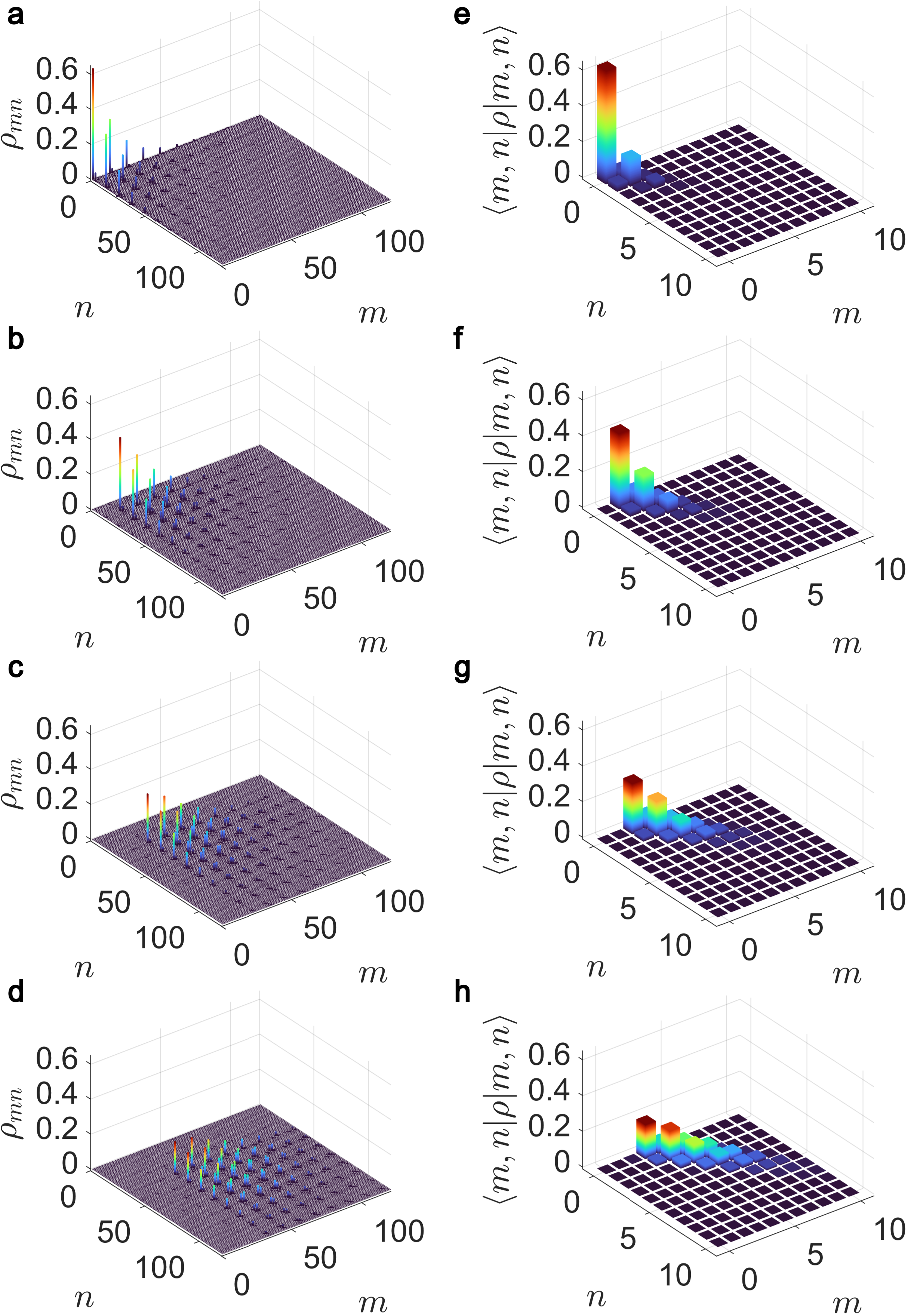}
\caption{Reconstructions of the joint density matrix in the photon number basis (\textbf{a}-\textbf{d}) for zero, one, two, and three added photons respectively, and the corresponding joint photon-number distributions (\textbf{e}-\textbf{h}). Only the real part of the density matrix is shown, because the imaginary part is about two orders of magnitude smaller than the real part and are all very close to zero. For \textbf{a}-\textbf{d}, the index $m$ on the X axis labels the basis vector $\ket{r}\ket{s}$, where $m = 11r+s+1$, with the indices $r$ and $s$ lying between 0 and 10. The index $n$ on the Y axis is defined similarly. For \textbf{e}-\textbf{h}, the labels $m$ and $n$ correspond to the photon numbers of each mode.} 
\label{F0}
\end{figure}

By inverting the homodyne and heterodyne measurements using maximum likelihood estimation, we can obtain a full tomographic reconstruction of the joint density matrix of the two optical modes (Fig.~\ref{F0}a-d), as well as the corresponding joint photon-number distribution (Fig.~\ref{F0}e-h). The effect of photon addition can be seen quite clearly in these images as a combination of translations in the photon number basis and a relative increase in the off-diagonal amplitudes with each added photon, with the latter indicating the increase in quantum correlations harboured by the state. From the density matrix, we can also obtain the purity of the states, which was 0.71, 0.55, 0.44, and 0.38 for zero, one, two and three added photons, respectively.

\begin{figure}[!h]
\centering\includegraphics[width=\textwidth]{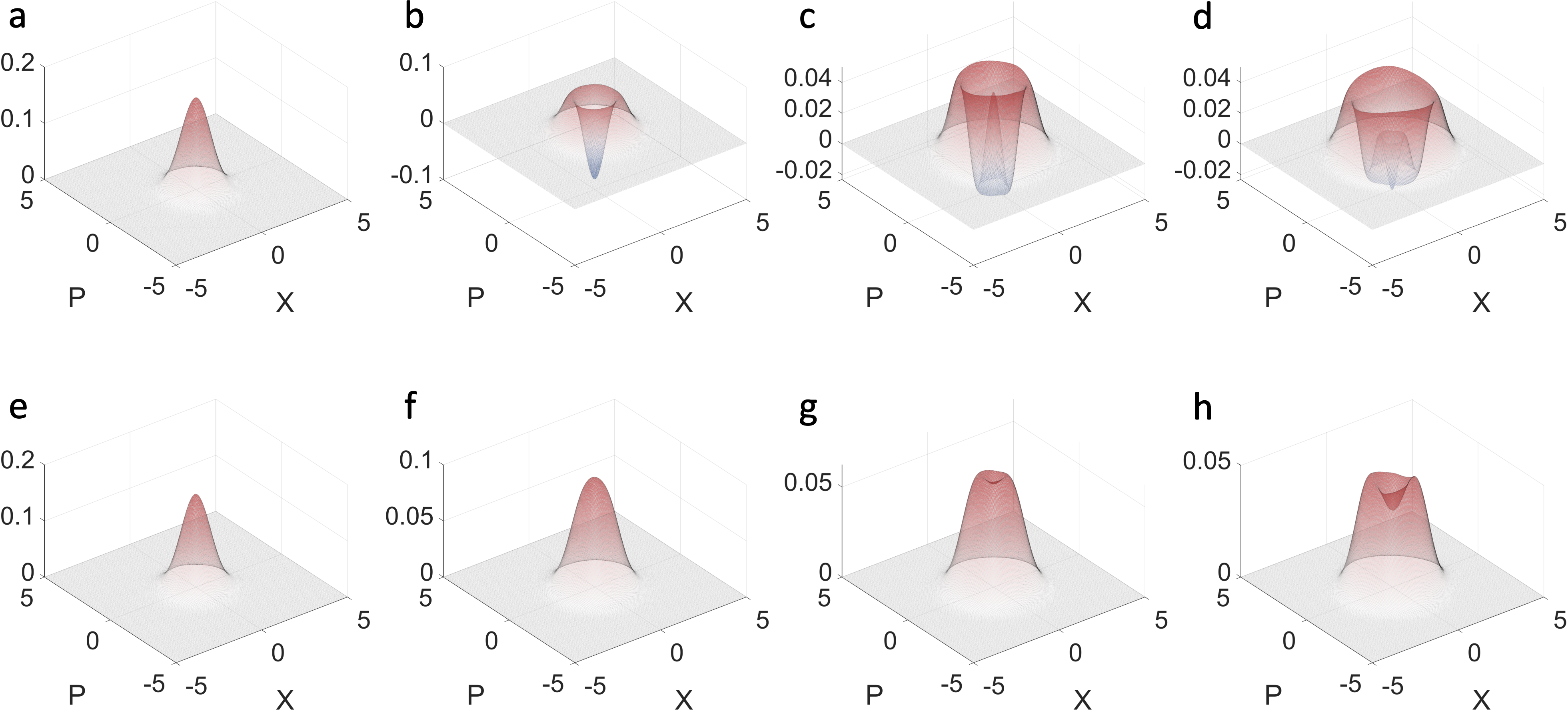}
\caption{Reconstructions of the local Wigner functions of photon-added two-mode-squeezed-vacuum states. \textbf{a}-\textbf{d} The mode to which photons are added, and \textbf{e}-\textbf{h} the mode which is untouched. The number of added photons increments by one from left to right, starting with the Gaussian state (\textbf{a} and \textbf{e}).}
\label{F1}
\end{figure}

Figure \ref{F1} shows local Wigner functions of the reconstructed state. Photon addition induces non-Gaussian features that look very similar to those observed in Fock states, i.e. radial oscillations with number equal to the number of added photons. The Wigner function of the photon-added mode is negative in alternating annular regions surrounding the origin and with the value at the origin oscillating between positive and negative; this negativity is not as large as that would be expected of ordinary Fock states, since we are adding photons to thermal states and not to the vacuum. The mode where no photons are added has a broadened distribution that remains positive everywhere but now sags at the origin; this is a consequence of the photon addition operation being probabilistic, and what we are seeing here is the Wigner function of the \textit{conditional} state heralded by a successful photon addition event.

\begin{figure}[!h]
\centering\includegraphics[width=\textwidth]{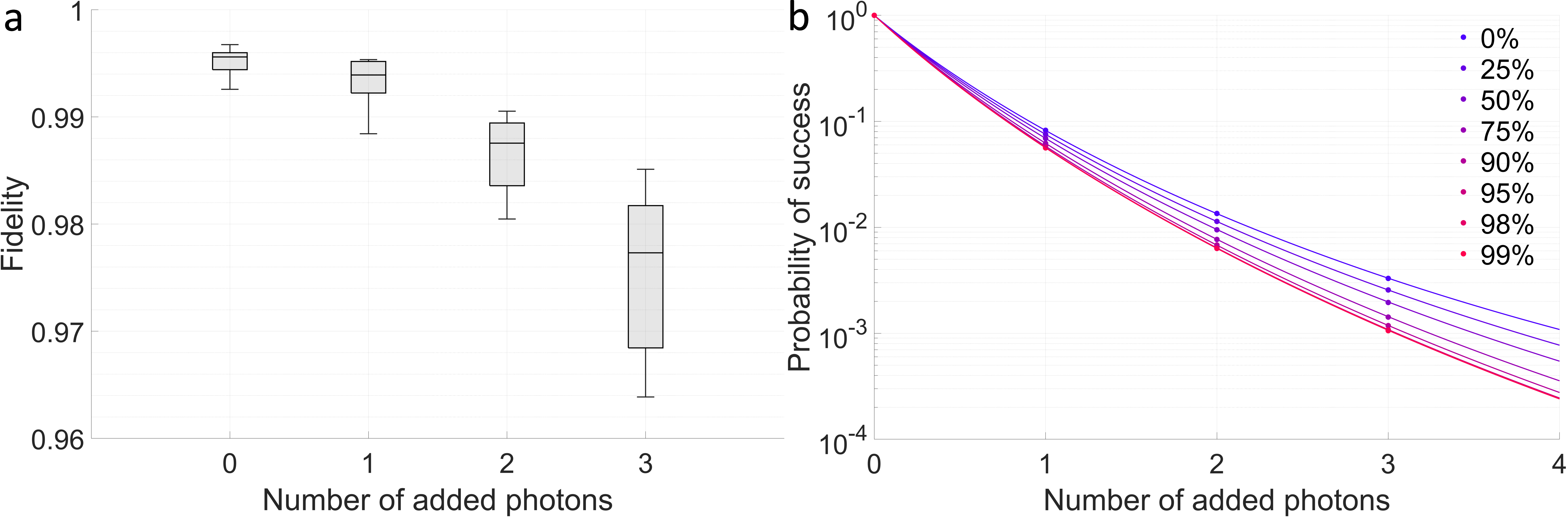}
\caption{\textbf{a} Reconstruction fidelity for photon-added two-mode-squeezed-vacuum states with various degrees of optical losses on the photon-added mode, represented using a box and whiskers plot. The variation of the fidelity across different degrees of optical loss exhibited no obvious trend, and have therefore been grouped into a single distribution for each number of added photons. \textbf{b} The probability of success. Solid points represent experimental data and solid lines represent the theoretical model (Eq.~\ref{eq:pos}).}
\label{F2}
\end{figure}

We observed that in all cases the quantum states were reconstructed with fidelities exceeding 96\% (Fig.~\ref{F2}a), with the Gaussian states (i.e. no photons added) reconstructed with fidelities exceeding 99\%. To calculate the fidelity, we take as the reference state $\rho'$ which corresponds to simulated photon addition on the measured Gaussian state at each loss level, and then calculate $(\Tr \sqrt{\sqrt{\rho} \rho' \sqrt{\rho}})^2$. The exceedingly high fidelities observed shows that the maximum likelihood estimation procedure described in the previous section is suitable for reconstructing the state. It is somewhat remarkable that only amplitude and phase measurements are required on the homodyning side in order to reconstruct the state.

The probability of successfully adding photons depends on how the photon addition is implemented as well as the specific state to which the photons are added. In our situation, the heterodyne outcomes $x+ip$ form a Gaussian probability distribution $\exp(-(x^2+p^2)/2\sigma^2)/(2\pi\sigma^2)$, so if we assume that a negligible number of outcomes lie outside the cut-off radius, then the probability of success will be:
\begin{equation}
\label{eq:pos}
    P_k(\sigma) = \frac{1}{2\pi\sigma^2}\int dx dp \frac{(x^2+p^2)^{2k}}{|\a_c|^{2k}} e^{-\frac{x^2+y^2}{2\sigma^2}} = \frac{2^k k! \sigma^{2k} }{|\a_c|^{2k}}.
\end{equation}
The formula is quite involved, but as a rule of thumb it corresponds to roughly one order of magnitude reduction for every photon added. Figure \ref{F2}b shows the observed probabilities which fit very well to this model. The closeness of the fit indicates that the effects of the cut-off is indeed negligible, and thus the photon addition operation is very close to the ideal.

\begin{figure}[!h]
\centering\includegraphics[width=\textwidth]{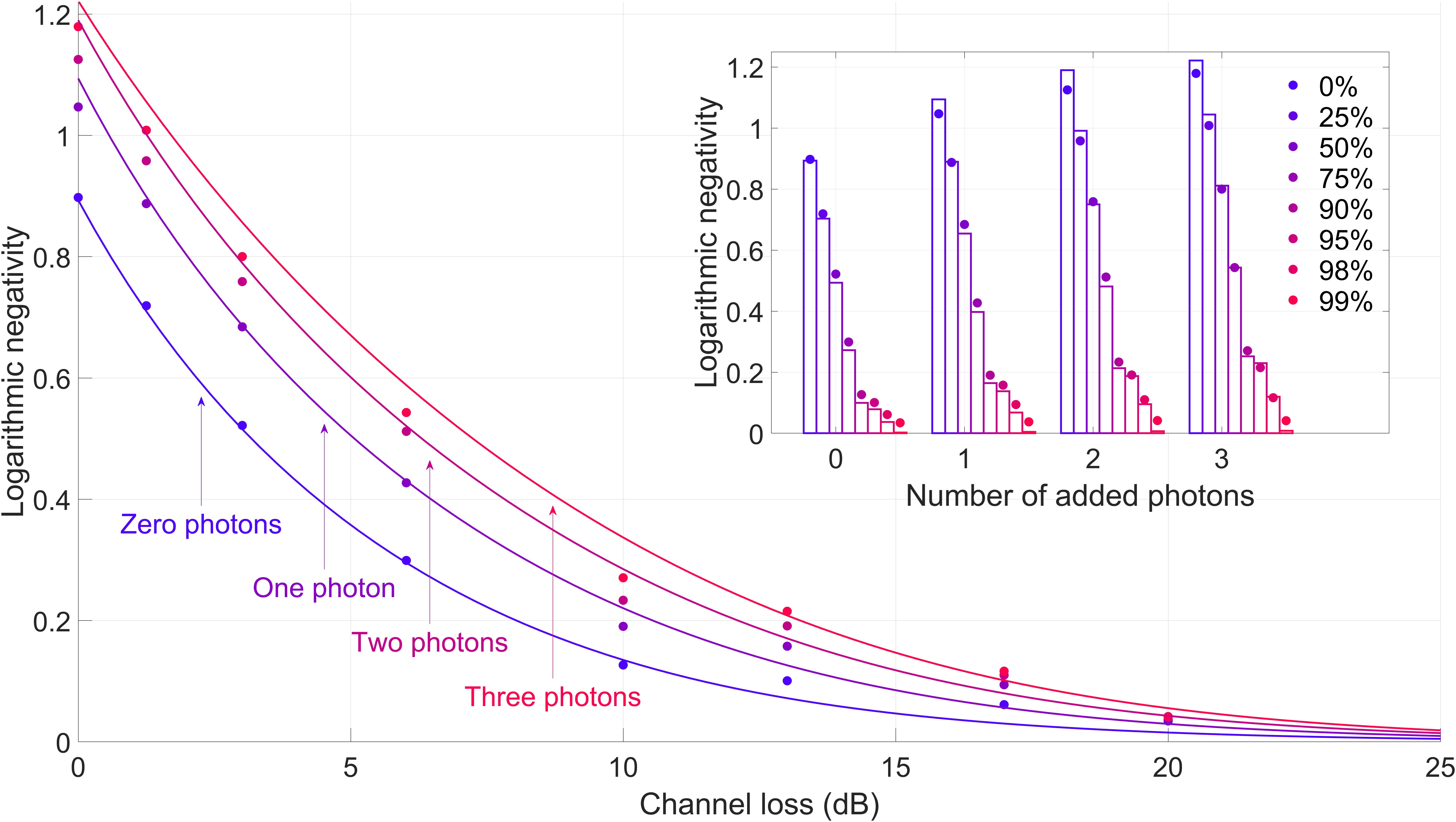}
\caption{Logarithmic negativity. Shown in the main figure is the dependence on channel loss, and in the inset the dependence on the number of added photons. Solid points represent experimental data, and solid lines and bars represent theoretical values. In the main figure the theoretical model is derived by taking the two-mode-squeezed-vacuum state at zero optical losses, and simulating both the effects of optical loss (beamsplitter with varying transmissivity) and ideal photon addition; in the inset, the model takes as starting point the Gaussian entangled state at each level of loss, and applies only the photon addition operation.}
\label{F3}
\end{figure}

We measure the degree of entanglement using logarithmic negativity \cite{vidal},
\begin{equation}
    E_N = \log ||\rho^\Gamma||_1,
\end{equation}
where $\Gamma$ denotes partial transposition and $||\rho||_1 = \Tr\sqrt{\rho\rho^\dagger}$ is the trace norm. The logarithmic negativity was observed to increase with the number of photons added, and to decrease with increasing channel losses (Fig.~\ref{F3}); it was found to always be positive for all the data we had, even for 99\% of loss. There appears to be some statistical fluctuations in the data which is due in large part to experimental imperfections (additional optical loss and phase noise), and in smaller part to the finite number of samples used which limits the accuracy of the reconstruction. A small degree of bias could be found in the limit of large channel losses. Similar results were obtained when the postselection and maximum likelihood estimation was repeated for selected datasets.

\begin{figure}[!h]
\centering\includegraphics[width=\textwidth]{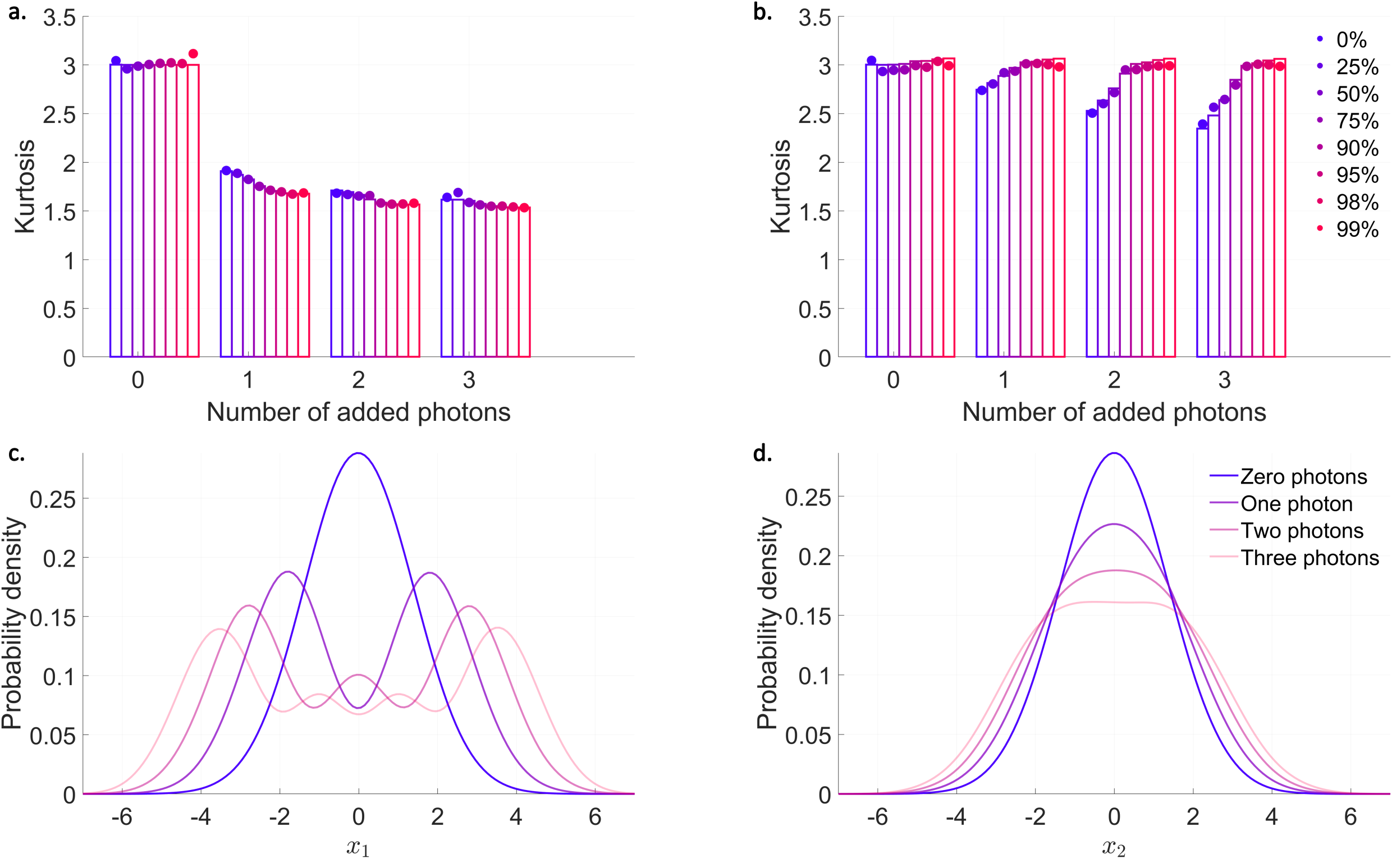}
\caption{\textbf{a} and \textbf{b} Kurtosis of local quadrature observables, solid points representing experimental data and hollow bars representing theoretical model. \textbf{c} and \textbf{d} Reconstructed marginals of the amplitude quadratures of each mode, for the case of zero channel losses. Legends are shared between \textbf{a} and \textbf{b}, and between \textbf{c} and \textbf{d}.} 
\label{F4}
\end{figure}

The kurtosis of a given quadrature observable is defined by the expression
\begin{equation}
    \Tr(\rho(\frac{x-\mu}{\sigma})^4),
\end{equation}
where $\mu = \Tr(\rho x)$ is the mean and $\sigma = \sqrt{\Tr(\rho x^2) - \Tr(\rho x)^2}$ is the standard deviation, and provides a quantitative description of the degree to which the state is non-Gaussian. It is equal to 3 for a Gaussian distribution, larger than 3 if the distribution has a shortened tail (and looks rather pointy), and less than 3 if the tail is longer (and looks rather blunt). It is found that after photon addition the kurtosis of all the local marginal distributions were less than 3, with the distributions broadened in a non-Gaussian manner (Fig.~\ref{F4}). The kurtosis of the photon-added mode falls sharply even just after one added photon, whereas the marginals of the freely-propagating mode remains relatively close to a Gaussian distribution, with only a slight correlated reduction.

\section{Quantum key distribution}


From the preceding discussions it is abundantly clear that photon-added two-mode-squeezed-vacuum states are very much non-Gaussian, and that it will be difficult to describe its properties using only the covariance matrix. To accurately account for the quantum correlations held by the state, we will need to carry out calculations without making any reference to the covariance matrix. The rate $K$ at which we can extract a pair of secret keys out of a shared entangled state is given by the formula
\begin{equation}
    K = I_{AB} - \chi_E,
    \label{eq:keyrate}
\end{equation}
where $I_{AB}$ denotes the mutual information between the two parties, Alice and Bob, given a certain measurement, and $\chi_E$ is the Holevo bound on the eavesdropper's (Eve) accessible information. Implicit in the expression above is the assumption of ideal reconciliation efficiency. If the state is assumed to be Gaussian, then both the mutual information and the Holevo information can be readily evaluated, depending analytically on only the covariance matrix associated with the state \cite{garcia-patron_thesis}. If the state is non-Gaussian, however, then no closed formula is known in general, and this is the case that we wish to deal with.

To evaluate the mutual information, we must first specify the measurements that are to be made. For simplicity we will use homodyne measurements on both sides since it often leads to the highest keyrates, though from a fundamental point of view it will not matter very much whether we use homodyne measurements or heterodyne measurements or a combination of homodyne and heterodyne, as the physical principles are the same. We therefore assume that Alice and Bob share a joint probability distribution associated with the measurements:
\begin{equation}
\label{eq:jpd}
    P(x,y) = \bra{x,y} \rho \ket{x, y}, 
\end{equation}
where $x$ and $y$ denote their respective homodyne measurement outcomes. Their locally observed probability distributions are given by
\begin{align}
    P_A(x) &= \int dy P(x,y), \\
    P_B(y) &= \int dx P(x,y),
\end{align}
and their mutual information by
\begin{equation}
    I_{AB} = \int dx dy P(x,y) \log \frac{P(x,y)}{P_A(x)P_B(y)}.
\end{equation}
The integrals can be evaluated numerically via Riemann sums if the distributions are not Gaussian.

Evaluation of Eve's information is slightly more tricky but follows a similar train of thought. Given an entangled state $\rho_{AB}$ shared between Alice and Bob, we first extend this state into a larger, pure, three-party entangled state $\rho_{ABE}$ that is now shared between Alice, Bob, and Eve. This can always be done, for given any density matrix $\rho = \sum_j p_j \ket{j}\bra{j}$, we can always form the pure state $\ket{\psi} = \sum_j \sqrt{p_j} \ket{j}\ket{j}$, which reduces to $\rho$ when one of the modes is traced out. It is not in general possible to specify uniquely such a diagonal representation of the density matrix, so there will correspondingly be many different ways of carrying out the purification. The specific choice of the purification will not matter however, because $\rho_{ABE}$ is a closed system, so that we can always rewrite the quantities of interest in terms of only the quantities determined by the quantum states of Alice and Bob.

Applying Holevo's theorem \cite{holevo}, Eve's accessible information with respect to Bob's measurement outcomes is bounded by
\begin{equation}
    I(B:E) \le \chi_E = S(\rho_E) - \int dy P_B(y) S(\rho_y),
\end{equation}
where $\rho_E = \Tr_{AB} \rho_{ABE}$ is Eve's local quantum state, $\rho_y$ is the conditional state held by Eve when Bob registers a particular outcome $y$,
\begin{equation}
    \rho_y = \Tr_{A}\bra{y}\rho_{ABE}\ket{y},
\end{equation}
and $S$ denotes the von Neumann entropy. The term $S(\rho_y)$ can be simplified by noting that $\bra{y}\rho_{ABE}\ket{y}$ is a pure state, so that we can swap the trace from that of Alice to that of Eve, and reduce the expression to only quantities that depend on the entangled state shared by Alice and Bob:
\begin{align}
    S(\rho_y) &= S(\Tr_{A}\bra{y}\rho_{ABE}\ket{y}) \\
    &= S(\Tr_{E}\bra{y}\rho_{ABE}\ket{y}) \\
    &= S(\bra{y}\rho_{AB}\ket{y}).
\end{align}
By the same reasoning we obtain also the von Neumann entropy of $\rho_E$, which is equal to the von Neumann entropy of Alice and Bob's shared entangled states, $S(\rho_E) = S(\rho_{AB})$. Thus we now have all the necessary information to calculate the keyrate (Eq.~\ref{eq:keyrate}). It is straightforward to check that this procedure indeed leads to the expected keyrates in the instances where the shared entangled state is Gaussian \cite{garcia-patron_thesis}.

\begin{figure}[!h]
\centering\includegraphics[width=\textwidth]{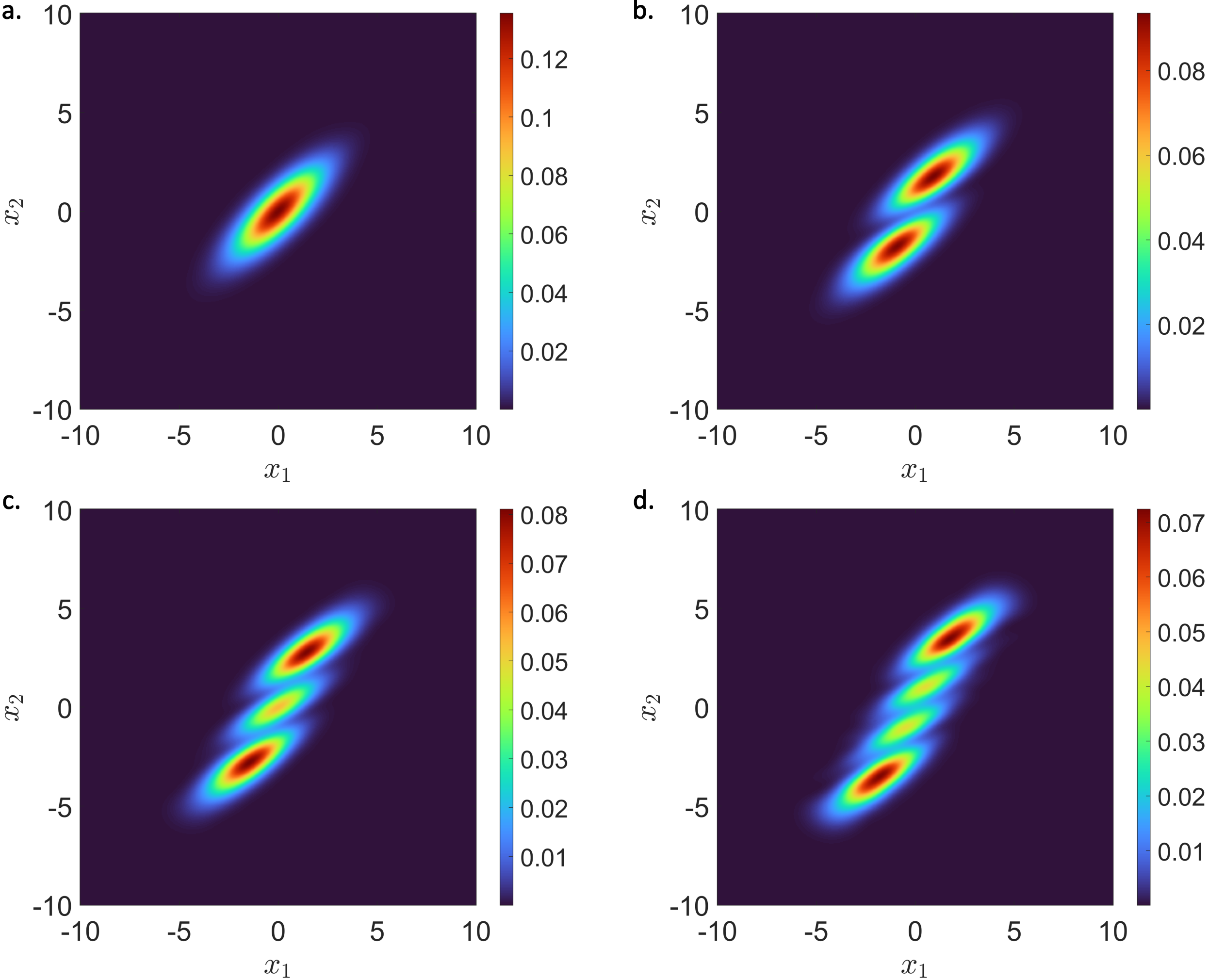}
\caption{Joint probability distributions between the amplitude quadratures of the two modes (zero channel loss). Number of photons added are incremented by one from \textbf{a} to \textbf{d}, with \textbf{a} being the initial Gaussian state. The distributions are calculated from the reconstructed density matrices using Eq.~\ref{eq:jpd}.}
\label{F5}
\end{figure}

Experimental reconstructions of the joint probability distributions are shown in Figure \ref{F5}, and the state could be seen to exhibit non-Gaussian correlations when at least one photon is added. The joint distributions for phase quadratures are identical except for a clockwise rotation by ninety degrees (i.e. negative correlations instead of positive), and have been suppressed on account of the similarity. These joint distributions are composed of multiple, parallel, elliptical regions of density, increasing by one in number with each photon added, and most densely concentrated at the two ends. There are therefore still correlations between the two observables, but certainly not of the same kind as that of two-mode-squeezed-vacuum states. In the limit when one beam is passed through very large optical losses, the lobes remain separated but now align along one axis, such that the correlations between the two variables are greatly diminished.

\begin{figure}[!h]
\centering\includegraphics[width=\textwidth]{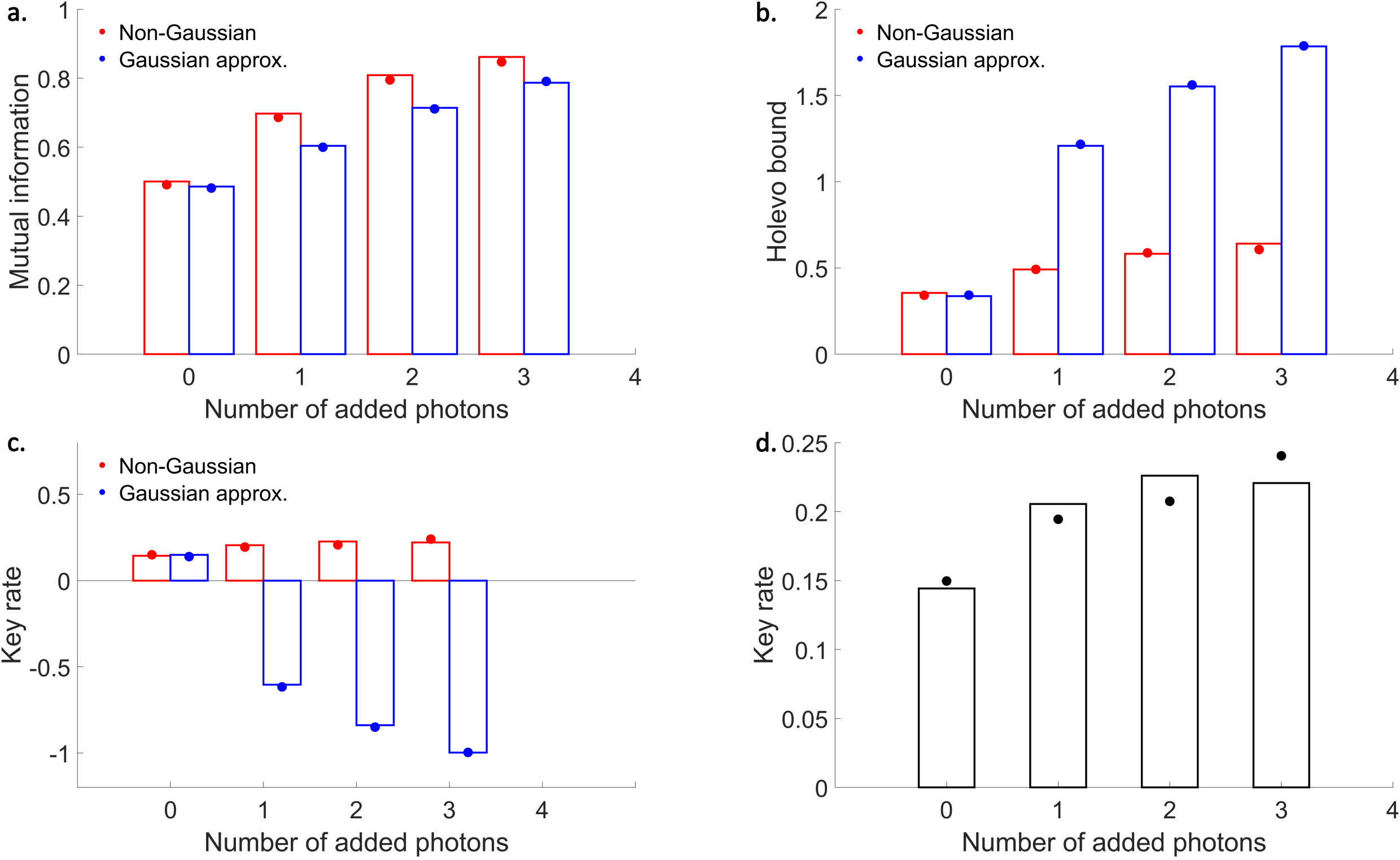}
\caption{Dependence of mutual information (\textbf{a}), Eve's accessible information (\textbf{b}), and the total keyrate (\textbf{c} and \textbf{d}) on the number of added photons, for the case of zero channel loss. Solid points represent experimental data, hollow bars represent theoretical model. For \textbf{a}, \textbf{b}, and \textbf{c}, data on the right (in blue) is obtained by making use of Gaussian extremity, while those on the left (in red) is without. \textbf{d} is a close-up image of the non-Gaussian keyrates in \textbf{c} with a better scale for comparing the relative increase between each added photon.}
\label{F6}
\end{figure}

Observations of the keyrates are shown in Figure \ref{F6}, decomposed into the positive part provided by the mutual information and the negative part due to Eve's information, and compared with the lower bounds given by the Gaussian extremity principle. We observed a small but marked increase in the mutual information when the Gaussian bound is not used (Fig.~\ref{F6}a), indicating that the photon addition does lead to stronger correlations between the amplitude (and phase) quadratures of the two beams of light, and that the Gaussian bound offers a reasonably good though imperfect approximation of the mutual information.

More strikingly, however, is the significant reduction in Eve's accessible information when the Gaussian bound is not used (Fig.~\ref{F6}b). The Holevo bound in the general case increases only very slightly, as opposed to the Gaussian bound which jumps suddenly by a factor of three when the first photon is added and which continues to increase steadily with further photon addition. Figure \ref{F6}c and d compares the results for keyrates obtained using the two methods, and the overestimation of the Holevo bound is found to be sufficient to drive the keyrate negative (i.e. no keys can be distributed securely) after the addition of the first photon, so that photon addition offers no advantage at all. In contrast, the full non-Gaussian keyrate increases with each additional added photon until the third added photon, where it appears that the experimental imperfections of the initially shared entangled states are such that no further advantage could be gotten. 

The above analyses pertain to the entangled state with no loss, but we observe similar results for any degree of losses. We note also that the keyrates shown in (Fig.~\ref{F6}c) do not incorporate the probabilistic overhead caused by photon addition, which will reduce the overall keyrate by a factor equal to the probability of success such that the keyrates for photon-added entangled states will generally fall below the keyrate for Gaussian states. What is interesting here, however, is that the number of secure key pairs we could extract from \textit{each} photon-added entangled state is greater than that from each Gaussian state.

\begin{figure}[!h]
\centering\includegraphics[width=\textwidth]{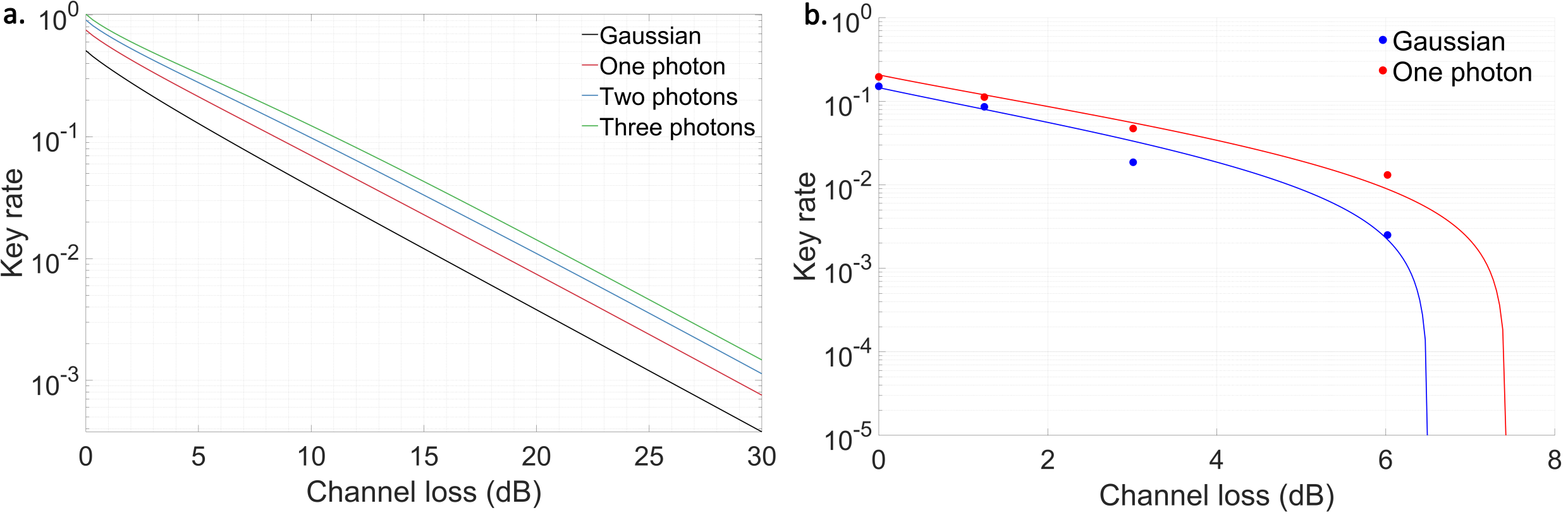}
\caption{Rate-loss dependence. \textbf{a} Simulated keyrates for ideal two-mode-squeezed-vacuum states with added photons (black: zero, red: one, blue: two, green: three). \textbf{b} Keyrates using the experimentally generated non-Gaussian correlations (blue: the Gaussian state, red: one photon added), with experimental data represented by solid points and theory by solid lines. Further addition of photons was found not to lead to any increase in keyrate, and clutters the figure and are therefore suppressed from this figure. The probability of success are not included in the keyrates of either figure. }
\label{F7}
\end{figure}

To understand the effect of long-distance optical transmission, we have performed simulations of the transmission of photon-added entangled states through various levels of optical attenuation (Fig.~\ref{F7}a). The simulation assumes the following:
\begin{enumerate}
    \item The initial entangled states are pure two mode squeezed vacuum states.
    \item No additional sources of inefficiencies other than optical attenuation.
    \item Both parties perform homodyne detection and implement reverse reconciliation.
    \item The probabilistic overhead is neglected (in order to assess the intrinsic keyrate of the state).
\end{enumerate}
Under these conditions, we find that the keyrate is positive no matter how many photons are added, increasing with the number of added photons at each level of channel losses. In the limit of very large channel losses, the keyrates converge to approximately the same scaling exponent (i.e. the same gradient on a logarithmic scale). No significant differences were observed if photon addition was put on the other mode without the optical attenuation apart from a slight reduction in the keyrates, however the keyrate would become negative at finite channel loss if photon addition came before the attenuation (on the same mode). It is not desirable to operate the protocol with forward information reconciliation for the same reason as normal Gaussian QKD \cite{grosshans3}.

We have observed that the addition of a single photon to the experimentally-generated two-mode-squeezed-vacuum states leads to greater intrinsic keyrates (Fig.~\ref{F7}b) compared to the Gaussian entangled state. Neither state allows for extraction of secret keys beyond a certain distance (owing to the experimental imperfections in preparing the two-mode-squeezed-vacuum state), which are 6.5 dB and 7.4 dB of channel losses respectively. Assuming 0.15 dB/km of optical attenuation for state-of-the-art optical fibres, these values correspond to $43$ km and $49$ km respectively, which is a 14\% increase in the distance. If the probabilistic overhead is included then most of the photon-added states would now give a lower keyrate compared to the Gaussian state, but the distance at which the protocol fails will remain unaffected. In other words, between the ranges of 43 to 49 km, only the photon-added entangled state would allow one to deliver a cryptographic key securely. Remarkably, the addition of the photon protects the system against decoherence, both passive (optical attenuation) and active (thermal noise on the entangled state).

We have also searched for signs that adding more photons might lead to further increase in keyrates, but that turned out not to be the case. This is most probably due to experimental imperfections, and suggest that photon added states are still fragile to some extent.

It is of interest to note that the keyrates of two and three added photons, as shown in Figure \ref{F7}a, has values greater than the upper bound on repeaterless communication: $K = -\log(1-T)$ \cite{pirandola}. The difference is quite small, around a maximum of 10\% increase for two photons added and around 50\% increase for three added photons. This means that if we were to somehow accumulate a stockpile of these entangled states, we could communicate securely at a higher rate than real-time communication using a maximally entangled EPR state; this could be possible using a quantum memory. There is of course no violation of the bound for our implementation of the photon addition operation however, not least because the probability of success is around 10\% when adding just one photon.

\begin{figure}[!h]
\centering\includegraphics[width=0.5\textwidth]{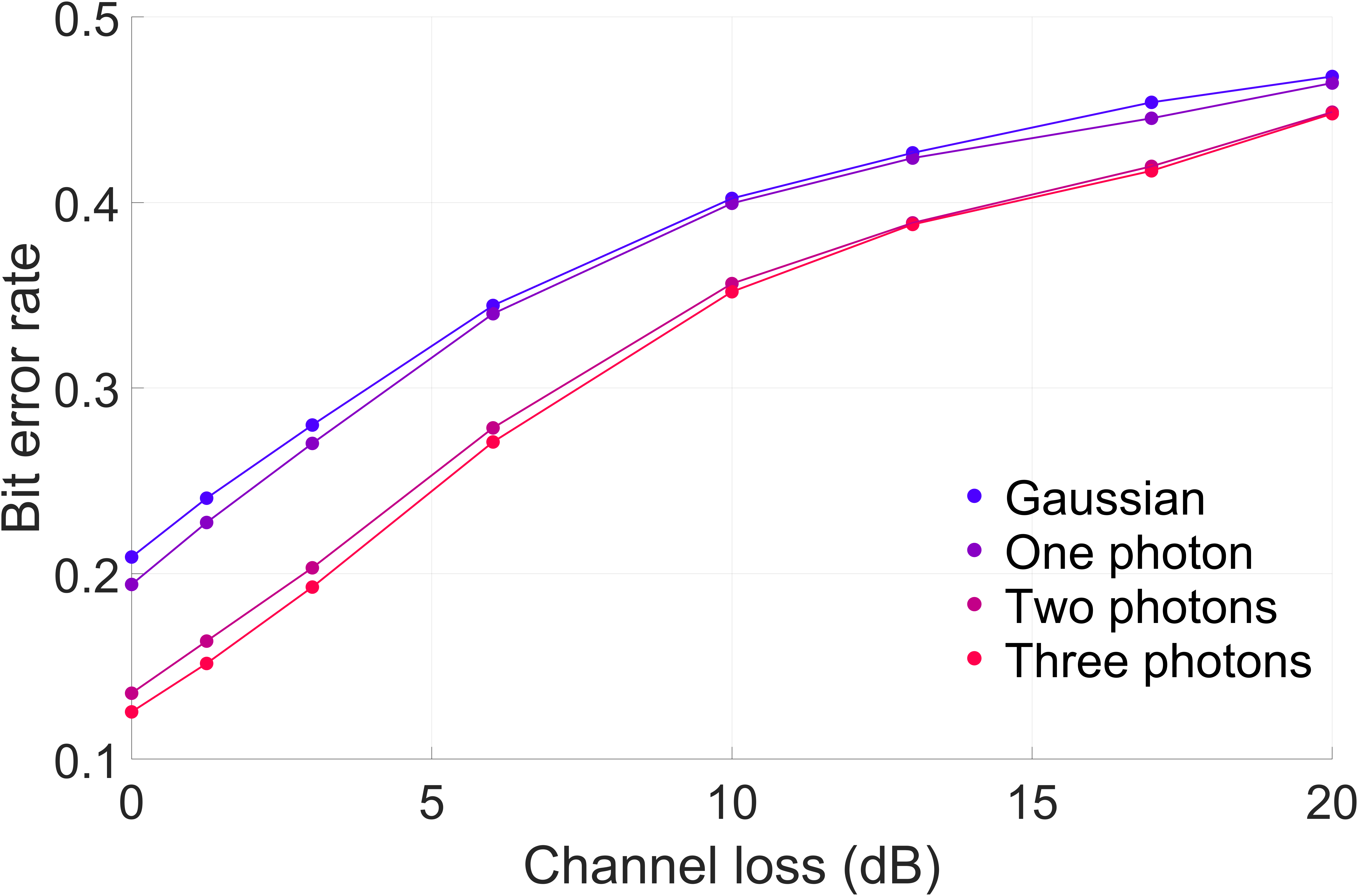}
\caption{Bit error rates for a binary encoding scheme, where the sign of homodyne outcomes is the bit and decoding achieved by maximum a posteriori estimation. Solid points represent experimental data and connecting lines act as a guide for the eye. Error rates are obtained through Monte Carlo simulation of one million samples of key exchanges using the experimentally determined density matrices; this number of sample size was found to be sufficient for convergence.}
\label{F8}
\end{figure}

To complete the analysis of this protocol, we need to specify also the method of encoding and decoding (i.e. how information reconciliation should be carried out). Since the keyrate is below 1 bit per use, we adopt a simple binary encoding using the sign of Alice's measurement outcome to encode one bit of information (positive values correspond to one, and negative values correspond to zero). For each of Bob's measurement outcome, he now has to guess what Alice's bit is, which can be achieved through maximum a posteriori estimation:
\begin{equation}
    \hat{X} = \argmax_x P(X=x|Y=y),
\end{equation}
and which would minimize the probability of decoding error.

Figure \ref{F8} shows the bit error rate of this encoding and decoding procedure when applied to our setup. It is found that the bit error rate for photon-added entangled states is at least as good as the Gaussian state, with marked reduction for each added photon. We have also carried out simulations to understand how the bit error rate behaves in a more general sense, and we found that the most important effect is the degree of squeezing present in the entangled state: higher squeezing levels lead to much lower bit error rates, which could be well below 10\% with added photons, indicating compatibility with current error correction protocols. The presence of experimental imperfections did not appear to affect the error rate very much, but of course it will increase with channel loss, which is also an issue limiting the range of Gaussian quantum key distribution. Taken together, these results indicate that the use of non-Gaussian entangled states does not present any difficulties at the stage of information reconciliation, and, indeed, appears to work even better compared to Gaussian states.

\section{Conclusion}

We have analysed in detail the non-Gaussian properties of photon-added two-mode-squeezed-vacuum states, and demonstrated how these features are advantageous for the secure distribution of cryptographic keys. This is quite the opposite to what the Gaussian extremity principle would have us believe, which is a protocol that has been compromised by no more than the mere illusion of an eavesdropper. The present scheme can be expanded and generalised in a number of ways, through other measurement schemes that could offer additional protection against noise \cite{weedbrook, garcia-patron_thesis}, more advanced encoding and decoding procedures that could simplify information reconciliation and reduce bit error rates \cite{van-assche, leverrier2}, and most certainly the use of different non-Gaussian entangled states produced through other combinations of photon creation and annihilation operations.

To obtain estimates of the key rate, we have made use of quantum state tomography. This is necessary due to the complexity of the quantum states that we are dealing with, and also possible owing to the use of homodyne and heterodyne detection. This situation might appear somewhat unusual given that partial measurements suffice to determine the security of many other QKD schemes, but it is not altogether alien. In many schemes operating under the assumption of Gaussian states, tomographic reconstruction is carried out implicitly---typically, we tend to speak of the quadrature means and covariance matrix and not of the density matrix itself, but we might as well be since there exists a simple relation between the density matrix and those quantities. In the general case of non-Gaussian states however, there is no simple way of parametrising the system, so it is most natural to work directly with the density matrix.

It is the primary focus of this paper to analyse in an intuitive way the fundamental effects that non-Gaussian entangled states have on quantum key distribution, thus in obtaining the keyrates we have made the simplifying assumption of asymptotic rates. Under this assumption, the homodyne measurements from which the keys would be extracted provide also sufficient information to determine the quantum state shared by the sender and the receiver. With a sufficiently large number of measurements, it is then possible to know precisely the ideal keyrate, as indicated by the small errors associated with the reconstruction fidelities.
      
For the same reason, we have not considered other potential attacks an eavesdropper could make on such non-Gaussian entangled states, since we are effectively assuming that the reconstructed state (which is a mixed state, owing to channel losses as well as some experimental imperfections) already incorporates all the effects of the eavesdropper. Had the eavesdropper carried out a different attack that resulted in quantum states deviating from those observed in this paper, the conclusions drawn in this paper would not strictly apply, but the same methods developed here would of course remain applicable to the study of such systems.
      
All of this is not to say, however, that the effect of a finite number of keys cannot be studied and incorporated into the protocol, and assumptions relaxed as to what the eavesdropper may or may not do. On the contrary, such questions lead to interesting further directions of research. One possibility is to find the worst possible state consistent with the observed data using alternative methods such as semidefinite programming \cite{lin, winick}, and bound the keyrate in this way. A more direct generalisation of the techniques employed in this paper is to use Bayesian analysis to both reconstruct the quantum state and bound the error of any quantity that depends on this state. The development of algorithms for fast Bayesian inference \cite{kim} and the application of these algorithms to the estimation of quantum states \cite{chapman} is a rapidly growing research area. By using such methods that provide bounds on keyrates rather than point-estimates, it will also be possible to account for the statistical uncertainties that arise due to the finite number of measurements, and to perform parameter estimation with a much smaller number of samples.

Finally, while we have made use of measurements and postselection in order to generate the non-Gaussian correlations, it is important to us that these correlations could in principle have been obtained from a physical entangled state. It is only under such conditions that our results would have any bearing on the general problem of communication using quantum repeaters. With respect to the possibility that postselection need not lead to outcomes always representing physical states, we refer the reader to an upcoming paper by Erk{\i}l{\i}\c{c} et al. \cite{erkilic}, where the case of arbitrary postselection for quantum key distribution with coherent states is treated, and which would be of interest in the context of direct quantum communication between trusted parties.

\enlargethispage{20pt}

\acknowledgements{This work was supported by the Australian Research Council (ARC) under the Centre of Excellence for Quantum Computation and Communication Technology (Grant No. CE170100012). We thank Helen Chrzanowski for making available the entanglement data on which our analysis is based. One of us (H.J.) wishes to acknowledge the use of computational resources at the Max Planck Institute for Multidisciplinary Sciences and at the High Performance Computing cluster hosted by the University of Göttingen, and expresses his gratitude to Prof. Claus Ropers for permission to complete this manuscript whilst employed in Göttingen.}


\vskip2pc

\bibliography{ref}

\nocite{jeng2}

\end{document}